\documentclass[twocolumn,aps,prl,floatfix,superscriptaddress]{revtex4-1}

\usepackage{hyperref}
\usepackage[utf8]{inputenc}

\usepackage{graphicx}
\usepackage{dcolumn}
\usepackage{amsmath}
\usepackage{supertabular}
\usepackage{multirow}

\usepackage{color}
\usepackage{cancel}
\usepackage{ulem}
\usepackage{array}
\usepackage{tabularx}

\usepackage{amsmath}
\usepackage{amssymb}
\usepackage{xcolor}
\definecolor{myOrange}{rgb}{1,0.5,0.}
\definecolor{myGreen}{rgb}{0.0,0.6,0.1}
 
\newcommand{\removetext}[1]{}

\newcommand{\aaa}{\ensuremath{\mathrm{A}+\mathrm{A}}}
\newcommand{\AuAu}{\ensuremath{\mathrm{Au}+\mathrm{Au}}}
\newcommand{\PbPb}{\ensuremath{\mathrm{Pb}+\mathrm{Pb}}}

\newcommand{\pp}{\ensuremath{{p+p}}} 

\newcommand{\sqrtsNN}{\ensuremath{\sqrt{s_\mathrm{NN}}}}

\newcommand{\pizero}{\ensuremath{\pi^0}}

\newcommand{\gammadir}{\ensuremath{\gamma_\mathrm{dir}}}
\newcommand{\gammarich}{\ensuremath{\gamma_\mathrm{rich}}}

\newcommand{\pT}{\ensuremath{p_\mathrm{T}}}

\newcommand{\pTtrack}{\ensuremath{p_\mathrm{T,track}}}

\newcommand{\pTjet}{\ensuremath{p_\mathrm{T,jet}}}

\newcommand{\Ajet}{\ensuremath{A_\mathrm{jet}}}

\newcommand{\ET}{\ensuremath{E_\mathrm{T}}}
\newcommand{\ETtrig}{\ensuremath{E_\mathrm{T}^\mathrm{trig}}}

\newcommand{\pTtrig}{\ensuremath{p_\mathrm{T}^\mathrm{trig}}}
\newcommand{\pTjetch}{\ensuremath{p_\mathrm{ T,jet}^\mathrm{ch}}}

\newcommand{\gev}{\ensuremath{\mathrm{GeV/}c}}

\newcommand{\antikT}{\ensuremath{\mathrm{anti-}k_\mathrm{T}}}

\newcommand{\rr}{\ensuremath{R}}

\newcommand{\dphi}{\ensuremath{\Delta\varphi}}

\newcommand{\Ntrig}{\ensuremath{N_\mathrm{trig}}}

\newcommand{\AAtoTrig}{\ensuremath{\mathrm{A+A}\rightarrow\mathrm{trig}}}
\newcommand{\AAtoTrigjet}{\ensuremath{\mathrm{A+A}\rightarrow\mathrm{trig+jet}}}
\newcommand{\dNjetdpT}{\ensuremath{\frac{{\rm d}N_\mathrm{jet}}{\mathrm{d}\pTjetch}}}

\newcommand{\Rbrtwofive}{\ensuremath{\mathfrak{R}^{0.2/0.5}}}

\newcommand{\IAA}{\ensuremath{{I}_\mathrm{AA}}}

\newcommand{\Compton}{\ensuremath{qg\rightarrow{\gamma}{q}}}
\newcommand{\Annihilation}{\ensuremath{q\bar{q}\rightarrow\gamma{g}}}

\begin{document}


\title{Measurement of in-medium jet modification using direct photon+jet and $\pi^0$+jet correlations in {\it p+p} and central Au+Au collisions at $\sqrt{s_\mathrm{NN}}=200$ GeV}

\bigskip



\affiliation{Academia Sinica}
\affiliation{Abilene Christian University, Abilene, Texas   79699}
\affiliation{AGH University of Krakow, FPACS, Cracow 30-059, Poland}
\affiliation{Argonne National Laboratory, Argonne, Illinois 60439}
\affiliation{American University in Cairo, New Cairo 11835, Egypt}
\affiliation{Ball State University, Muncie, Indiana, 47306}
\affiliation{Brookhaven National Laboratory, Upton, New York 11973}
\affiliation{University of Calabria \& INFN-Cosenza, Rende 87036, Italy}
\affiliation{University of California, Berkeley, California 94720}
\affiliation{University of California, Davis, California 95616}
\affiliation{University of California, Los Angeles, California 90095}
\affiliation{University of California, Riverside, California 92521}
\affiliation{Central China Normal University, Wuhan, Hubei 430079 }
\affiliation{University of Illinois at Chicago, Chicago, Illinois 60607}
\affiliation{Chongqing University, Chongqing, 401331}
\affiliation{Creighton University, Omaha, Nebraska 68178}
\affiliation{Czech Technical University in Prague, FNSPE, Prague 115 19, Czech Republic}
\affiliation{Technische Universit\"at Darmstadt, Darmstadt 64289, Germany}
\affiliation{National Institute of Technology Durgapur, Durgapur - 713209, India}
\affiliation{ELTE E\"otv\"os Lor\'and University, Budapest, Hungary H-1117}
\affiliation{Frankfurt Institute for Advanced Studies FIAS, Frankfurt 60438, Germany}
\affiliation{Fudan University, Shanghai, 200433 }
\affiliation{Guangxi Normal University, Guilin, 541004}
\affiliation{University of Heidelberg, Heidelberg 69120, Germany }
\affiliation{University of Houston, Houston, Texas 77204}
\affiliation{Huzhou University, Huzhou, Zhejiang  313000}
\affiliation{Indian Institute of Science Education and Research (IISER), Berhampur 760010 , India}
\affiliation{Indian Institute of Science Education and Research (IISER) Tirupati, Tirupati 517507, India}
\affiliation{Indian Institute Technology, Patna, Bihar 801106, India}
\affiliation{Indiana University, Bloomington, Indiana 47408}
\affiliation{Institute of Modern Physics, Chinese Academy of Sciences, Lanzhou, Gansu 730000 }
\affiliation{University of Jammu, Jammu 180001, India}
\affiliation{Kent State University, Kent, Ohio 44242}
\affiliation{University of Kentucky, Lexington, Kentucky 40506-0055}
\affiliation{Lanzhou University,Lanzhou, Gansu, 730000}
\affiliation{Lawrence Berkeley National Laboratory, Berkeley, California 94720}
\affiliation{Lehigh University, Bethlehem, Pennsylvania 18015}
\affiliation{Max-Planck-Institut f\"ur Physik, Munich 80805, Germany}
\affiliation{Michigan State University, East Lansing, Michigan 48824}
\affiliation{National Institute of Science Education and Research, HBNI, Jatni 752050, India}
\affiliation{National Cheng Kung University, Tainan 70101 }
\affiliation{Nuclear Physics Institute of the CAS, Rez 250 68, Czech Republic}
\affiliation{The Ohio State University, Columbus, Ohio 43210}
\affiliation{Panjab University, Chandigarh 160014, India}
\affiliation{Purdue University, West Lafayette, Indiana 47907}
\affiliation{Rice University, Houston, Texas 77251}
\affiliation{Rutgers University, Piscataway, New Jersey 08854}
\affiliation{University of Science and Technology of China, Hefei, Anhui 230026}
\affiliation{South China Normal University, Guangzhou, Guangdong 510631}
\affiliation{Sejong University, Seoul, 05006, South Korea}
\affiliation{Shandong University, Qingdao, Shandong 266237}
\affiliation{Shanghai Institute of Applied Physics, Chinese Academy of Sciences, Shanghai 201800}
\affiliation{Southern Connecticut State University, New Haven, Connecticut 06515}
\affiliation{State University of New York, Stony Brook, New York 11794}
\affiliation{Instituto de Alta Investigaci\'on, Universidad de Tarapac\'a, Arica 1000000, Chile}
\affiliation{Temple University, Philadelphia, Pennsylvania 19122}
\affiliation{Texas A\&M University, College Station, Texas 77843}
\affiliation{University of Texas, Austin, Texas 78712}
\affiliation{Tsinghua University, Beijing 100084}
\affiliation{University of Tsukuba, Tsukuba, Ibaraki 305-8571, Japan}
\affiliation{University of Chinese Academy of Sciences, Beijing, 101408}
\affiliation{United States Naval Academy, Annapolis, Maryland 21402}
\affiliation{Valparaiso University, Valparaiso, Indiana 46383}
\affiliation{Variable Energy Cyclotron Centre, Kolkata 700064, India}
\affiliation{Warsaw University of Technology, Warsaw 00-661, Poland}
\affiliation{Wayne State University, Detroit, Michigan 48201}
\affiliation{Wuhan University of Science and Technology, Wuhan, Hubei 430065}
\affiliation{Yale University, New Haven, Connecticut 06520}

\author{B.~E.~Aboona}\affiliation{Texas A\&M University, College Station, Texas 77843}
\author{J.~Adam}\affiliation{Czech Technical University in Prague, FNSPE, Prague 115 19, Czech Republic}
\author{L.~Adamczyk}\affiliation{AGH University of Krakow, FPACS, Cracow 30-059, Poland}
\author{I.~Aggarwal}\affiliation{Panjab University, Chandigarh 160014, India}
\author{M.~M.~Aggarwal}\affiliation{Panjab University, Chandigarh 160014, India}
\author{Z.~Ahammed}\affiliation{Variable Energy Cyclotron Centre, Kolkata 700064, India}
\author{D. M. Anderson}\affiliation{Texas A\&M University, College Station, Texas 77843}
\author{E.~C.~Aschenauer}\affiliation{Brookhaven National Laboratory, Upton, New York 11973}
\author{S.~Aslam}\affiliation{Indian Institute Technology, Patna, Bihar 801106, India}
\author{J.~Atchison}\affiliation{Abilene Christian University, Abilene, Texas   79699}
% ...

\collaboration{STAR Collaboration}\noaffiliation


\date{ \today}



 
\begin{abstract}

The STAR Collaboration presents measurements of the semi-inclusive distribution of charged-particle jets recoiling from energetic direct-photon (\gammadir) and neutral-pion (\pizero) triggers in \pp\ and central \AuAu\ collisions at $\sqrtsNN=200$ GeV over a broad kinematic range, for jet resolution parameters $\rr=0.2$ and 0.5. Medium-induced jet yield suppression is observed to be larger for $\rr=0.2$ than for 0.5, reflecting the angular range of jet energy redistribution due to quenching. The predictions of model calculations incorporating jet quenching are not fully consistent with the observations. These results provide new insight into the physical origins of jet quenching.

\end{abstract}

\maketitle
 

{\it Introduction -} Strongly-interacting matter at high density and temperature forms a state of deconfined quarks and gluons called the Quark-Gluon Plasma (QGP)~\cite{Collins:1974ky,Shuryak:1977ut}. The QGP filled the early universe a few microseconds after the Big Bang, and is recreated today using energetic collisions of heavy atomic nuclei at the Relativistic Heavy Ion Collider (RHIC) and the Large Hadron Collider (LHC) ~\cite{Busza:2018rrf,Harris:2023tti}.

Jets are correlated sprays of hadrons arising from the fragmentation of energetic quarks and gluons generated in hard (high momentum transfer) partonic interactions~\cite{Abelev:2006uq,Adamczyk:2016okk,Abelev:2013fn,Aad:2014vwa,Khachatryan:2016mlc}. In heavy-ion collisions, jets propagate in the QGP and interact with it (``jet quenching'')~\cite{Bjorken:1982tu,Wang:1994fx,Baier:1998yf,Gyulassy:2000fs,Majumder:2010qh,Cunqueiro:2021wls,Apolinario:2022vzg}. Comparison of measurements and calculations of in-medium jet modification probes the structure and dynamics of the QGP~\cite{Majumder:2010qh,Cunqueiro:2021wls,Apolinario:2022vzg}. Reconstructed jet measurements in heavy-ion collisions are challenging, however, due to the large background from uncorrelated processes~\cite{Cunqueiro:2021wls}, and the initial observation of jet quenching was based on inclusive production and correlations of high transverse momentum (high-\pT) hadrons~\cite{Adler:2002xw,Adler:2002tq,Adams:2003kv,Adams:2006yt,Adamczyk:2013jei,Adcox:2001jp,Adare:2012wg,Adare:2010ry,Aamodt:2011vg,Abelev:2012hxa,ALICE:2016gso,CMS:2012aa,Chatrchyan:2012wg}. 

High-\pT\ hadron measurements are sensitive primarily  to medium-induced energy loss of the most energetic branch of the jet shower~\cite{Majumder:2010qh}. Deeper exploration of jet quenching mechanisms at the partonic level requires measurements of the full jet shower and its in-medium modification, using reconstructed jets ~\cite{Abelev:2013kqa,Adam:2015ewa,Acharya:2019jyg,Aad:2014bxa,Khachatryan:2016jfl,Adam:2015doa,Aad:2010bu,Chatrchyan:2012nia, Chatrchyan:2012gt,Sirunyan:2017qhf,Sirunyan:2018qec,Acharya:2017goa,Acharya:2019djg,ALICE:2023jye, ALICE:2023qve,Sirunyan:2017bsd,STAR:2016dfv,Adamczyk:2017yhe,Adam:2020wen,Aaboud:2018anc,CMS:2018jco,ATLAS:2023iad}. Of special interest is the correlation between a neutral vector boson  (direct photon \gammadir, $Z$) and a recoil jet. Vector bosons are colorless and do not interact significantly with the QGP~\cite{PHENIX:2012jbv,ALICE:2015xmh,ATLAS:2015rlt,CMS:2020oen,
CMS:2011zfr,ATLAS:2012qdj}, providing an unmodified reference for measuring jet quenching~\cite{Wang:1996yh}. At leading perturbative order, the dominant channel for \gammadir+jet production at RHIC energies is QCD Compton scattering, \Compton, in which the direct photon and the quark jet are balanced in \pT. Next-to-leading-order (NLO) effects modify this simple picture~\cite{Dai:2012am}. Measurements of \gammadir+jet coincidence observables in \pp\ and \PbPb\ collisions at the LHC exhibit in-medium modification of recoil jets 
~\cite{Chatrchyan:2012gt,Sirunyan:2017qhf,Aaboud:2018anc,CMS:2018jco,ATLAS:2023iad}. $Z$-boson/\gammadir+hadron measurements have also been reported~\cite{STAR:2009ojv,STAR:2016jdz,PHENIX:2009cvn,PHENIX:2010vgy,PHENIX:2012aba,PHENIX:2020alr,ATLAS:2020wmg,CMS:2021otx}. To date, no \gammadir-triggered correlation measurements with reconstructed jets have been reported at RHIC.

Measurements of high-\pT\ hadron+jet correlations~\cite{Adam:2015doa,Adamczyk:2017yhe,ALICE:2023jye, ALICE:2023qve} are likewise of interest. High-\pT\ hadrons are leading jet fragments, arising from a different mixture of partonic processes than direct photons, and are expected to have a larger fraction initiated by gluons and larger in-medium path length~\cite{STAR:2023ksv,He:2024rcv}. The comparison of similar measurements of \gammadir\ and hadron-triggered correlations can therefore provide novel constraints on the flavor and path-length dependence of jet quenching.


Current calorimetric \gammadir+jet measurements in central \PbPb\ collisions at the LHC are restricted to jet resolution parameter $\rr=0.3$ or 0.4 and $\pTjet>30$ or 40 \gev~\cite{Chatrchyan:2012gt,Sirunyan:2017qhf,Aaboud:2018anc,CMS:2018jco,ATLAS:2023iad}. This restricted $\pTjet$ range covers only the high tail of the $\pTjet$ distribution at RHIC. Jets reconstructed from charged particles only (also called ``track jets'') have greater instrumental precision than calorimetric jets, and provide the highest-precision jet substructure measurements at the 
LHC~\cite{Lee:2023xzv}. Charged-particle jets are calculable in pQCD using non-perturbative Track Functions extracted from data~\cite{Lee:2023xzv}. Monte Carlo generators at NLO accuracy agree well with charged-particle jet distributions measured in \pp\ collisions for inclusive production~\cite{ALICE:2023waz} and correlations, including low \pTjetch ($\sim10$ \gev)~\cite{ALICE:2023qve,ALICE:2023jye}. Charged-particle jets have been used to measure coincidence observables incorporating jets in heavy-ion collisions at RHIC and the LHC~\cite{Adam:2015doa,Adamczyk:2017yhe,ALICE:2023jye, ALICE:2023qve} to cover broader phase space in \pTjet\ and \rr\ than has been achieved by calorimeter-based measurements~\cite{Chatrchyan:2012gt,Sirunyan:2017qhf,Aaboud:2018anc,ATLAS:2023iad}. 

In this Letter and companion article~\cite{STAR:2023ksv}, the STAR Collaboration presents the first measurement of jet quenching using \gammadir+jet and \pizero+jet correlations in \pp\ collisions and central (0--15\%) \AuAu\ collisions at $\sqrtsNN=200$ GeV. The analysis is based on previous developments for \gammadir/\pizero\ discrimination~\cite{STAR:2009ojv,STAR:2016jdz} and for statistical correction of the complex jet background in central \AuAu\ collisions with mixed events (ME)~\cite{Adamczyk:2017yhe}. Charged--particle jets are reconstructed using the \antikT\ algorithm~\cite{Cacciari:2008gp} with $\rr=0.2$ and 0.5. Semi--inclusive recoil--jet distributions for \gammadir\ (transverse energy $15<\ETtrig<20$ GeV) and \pizero\ triggers ($11<\pTtrig<15$ \gev) are reported here. Theoretical models incorporating jet quenching are compared to the data, and constraints due to these comparisons on the physical mechanisms underlying jet quenching are discussed. Details of the analysis and models, and analysis for other trigger kinematics, are provided in~\cite{STAR:2023ksv}.


{\it Dataset and analysis -} The STAR detector is described in~\cite{Ackermann:2002ad}. The \pp\ and \AuAu\ datasets were recorded during the 2009 and 2014 RHIC runs, respectively. Online event selection utilized a trigger based on the STAR Barrel Electromagnetic Calorimeter (BEMC)~\cite{Beddo:2002zx} requiring transverse energy (\ET) in a single calorimeter tower greater than 4.2 GeV in \pp\ and 5.9 GeV in \AuAu\ collisions; and total \ET\ in the trigger and up to two adjacent towers greater than 7.44 GeV, in both \pp\ and \AuAu\ collisions. Event--selection cuts are applied to suppress pileup and to remove bad runs. After all event selection cuts, the integrated luminosity accepted for the analysis is 3.9 nb$^{-1}$ for \AuAu\ collisions and 23 pb$^{-1}$ for \pp\ collisions. Offline analysis follows the procedures described in 
\cite{STAR:2016jdz,Adamczyk:2017yhe}. 

Centrality in \AuAu\ collisions is determined using the uncorrected charged-particle multiplicity measured in pseudo-rapidity $|\eta|<0.5$~\cite{STAR:2018zdy}. Central \AuAu\ collisions in this analysis correspond to the 15\% highest-multiplicity events of the Minimum Bias (MB) population. 

High-\ET\ photons are generated by several sources: direct production from  partonic \Compton\ and \Annihilation\ processes; fragmentation from partonic bremsstrahlung; and hadronic decays, primarily of \pizero\ and $\eta$ mesons. Since the channel of interest for the analysis is direct production, the decay photon contribution is subtracted. The fragmentation photon contribution is suppressed but cannot be fully subtracted, as in all direct photon analyses.

Discrimination of high-\pT\ single-photon and \pizero-decay BEMC showers is based on the transverse-shower profile distribution measured by the Barrel Shower Maximum Detector~\cite{STAR:2016jdz}. The \pizero\ purity is $>95$\%. However, the single-photon population contains direct photons (\gammadir) with an admixture of fragmentation and decay photons (``gamma-rich'', \gammarich). The \gammadir\ fraction of the \gammarich\ population is measured by the rate of correlated charged tracks with $\pT>1.2$ \gev\ that are nearby in phase space, assuming that direct-photon triggers have zero correlated rate~\cite{STAR:2016jdz}. The \gammadir\ purity is \ET-dependent, varying from 67\% to 84\% for central \AuAu\ collisions and 43\% to 53\% for \pp\ collisions. 


Charged-particle tracks are reconstructed offline using signals from the STAR Time Projection Chamber (TPC)~\cite{Anderson:2003ur}. Jet reconstruction 
utilizes primary tracks, which include the event vertex in the momentum fit. The acceptance for primary tracks is $|\eta| <$ 1.0 and  $0.2<\pTtrack<30$ \gev, over the full azimuth. The tracking efficiency for primary tracks with $\pTtrack>2$ \gev\ is 72\% in central \AuAu\ collisions and 82\% in \pp\ collisions. Primary-track momentum resolution is approximately  $\sigma_{\pT}/\pT=0.01\cdot\pT/(\gev)$ in both systems.

Charged-particle jets are reconstructed from primary tracks by the \antikT~\cite{Cacciari:2008gp} algorithm with $\rr=0.2$ and 0.5, $E$-scheme recombination, and active ghost area 0.01~\cite{Cacciari:2011ma}. Jet candidates are accepted if their momentum-weighted centroid lies within $|\eta|<1.0-\rr$ and their area satisfies $\Ajet>0.05$ for $\rr=0.2$ and $\Ajet>0.65$ for $\rr=0.5$~\cite{Adamczyk:2017yhe}. The Jet Energy Resolution and Jet Energy Scale are similar to those in Ref.~\cite{Adamczyk:2017yhe}. The measured jet \pT\ is adjusted using the approximate background energy density $\rho$~\cite{Cacciari:2007fd}; this approximation is improved by the unfolding correction at a later analysis stage.



Events with a \gammarich\ or \pizero\ trigger are selected for the correlation analysis. The recoil-jet azimuthal acceptance is $\frac{3\pi}{4}<\dphi<\frac{5\pi}{4}$, where \dphi\ is the azimuthal separation between the trigger and jet centroid. All jet candidates satisfying this cut are accepted. 

The trigger-normalized semi-inclusive recoil-jet distribution is then tabulated. In the absence of uncorrelated background, the trigger-normalized distribution is equivalent to the ratio of perturbatively calculable cross sections~\cite{Adam:2015doa,Adamczyk:2017yhe}, 

\begin{equation}
\frac{1}{\Ntrig}\cdot\dNjetdpT\Bigg\vert_{\pTtrig}
= \left(
\frac{1}{\sigma^{\AAtoTrig}} \cdot
\frac{{\rm d}\sigma^{\AAtoTrigjet}}{\mathrm{d}\pTjetch}\right)
\Bigg\vert_{\pTtrig},
\label{eq:hJetDefinition}
\end{equation}

\noindent
where \Ntrig\ is the number of triggers, \dNjetdpT\ is the \pT-differential recoil-jet distribution, $\sigma^{\AAtoTrig}$ is the inclusive production cross section for the trigger,  and $\sigma^{\AAtoTrigjet}$ is the production cross section for the trigger and a recoil jet in the acceptance. 

The reconstructed jet population in central \AuAu\ collisions includes a large uncorrelated jet yield, which is corrected at the level of ensemble-averaged distributions by the ME procedure~\cite{Adamczyk:2017yhe}. After subtraction of the uncorrelated jet yield distribution, correction for \pT-smearing due to instrumental effects and residual background fluctuations is carried out using unfolding~\cite{Hocker:1995kb,DAgostini:1994fjx}. The corrected distribution for \gammadir\ triggers 
is determined from the linear combination of \gammarich\ and \pizero-triggered distributions, accounting for the \gammadir\ purity~\cite{STAR:2016jdz}.


The largest systematic uncertainties in central \AuAu\ collisions arise from the uncorrelated background yield correction; choice of unfolding algorithm, regularization, and prior; response matrix; tracking efficiency; and the \gammadir\ purity.  In \pp\ collisions, the largest systematic uncertainties are due to the tracking efficiency, unfolding, and the \gammadir\ purity. The cumulative uncertainty is the quadrature sum of all uncertainty components, and is shown in all figures as a continuous band to indicate large off-diagonal covariance~\cite{Adamczyk:2017yhe,Adam:2020wen}.

\begin{figure}[htb!]
\centering 	 	
\includegraphics[width=0.5\textwidth]{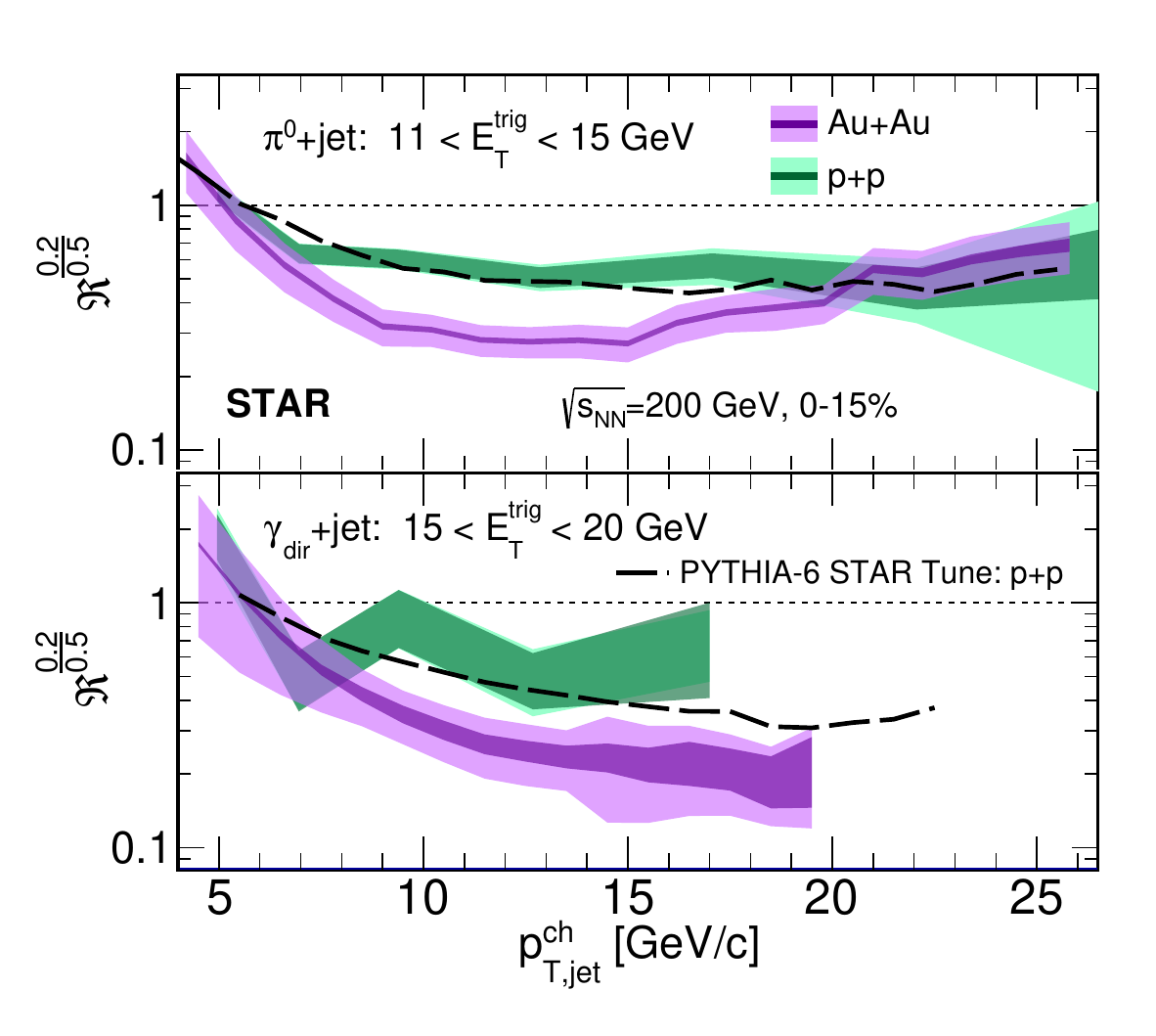}
\caption{(Color online) Ratio \Rbrtwofive\ of semi-inclusive recoil jet distributions in \pp\ and central \AuAu\ collisions~\cite{STAR:2023ksv} for \pizero\ (upper) and \gammadir\ (lower) triggers. Uncertainty bands account for correlated uncertainties in numerator and denominator. The dashed line shows a calculation for \pp\ collisions using PYTHIA-6 STAR tune.} 
\label{Fig:Rratio} 
\end{figure}

The distribution of \ETtrig\ is not corrected for BEMC energy resolution. The procedure to account for this resolution in theoretical calculations, to enable accurate comparison to data, is specified in Ref.~\cite{STAR:2023ksv}.  The PYTHIA calculations shown in Figs.~\ref{Fig:Rratio} and~\ref{Fig:RratioModels} account for the trigger-energy resolution, although the effects largely cancel in the ratios presented here.

\begin{figure}[htb!]
\centering
\includegraphics[width=0.55\textwidth]{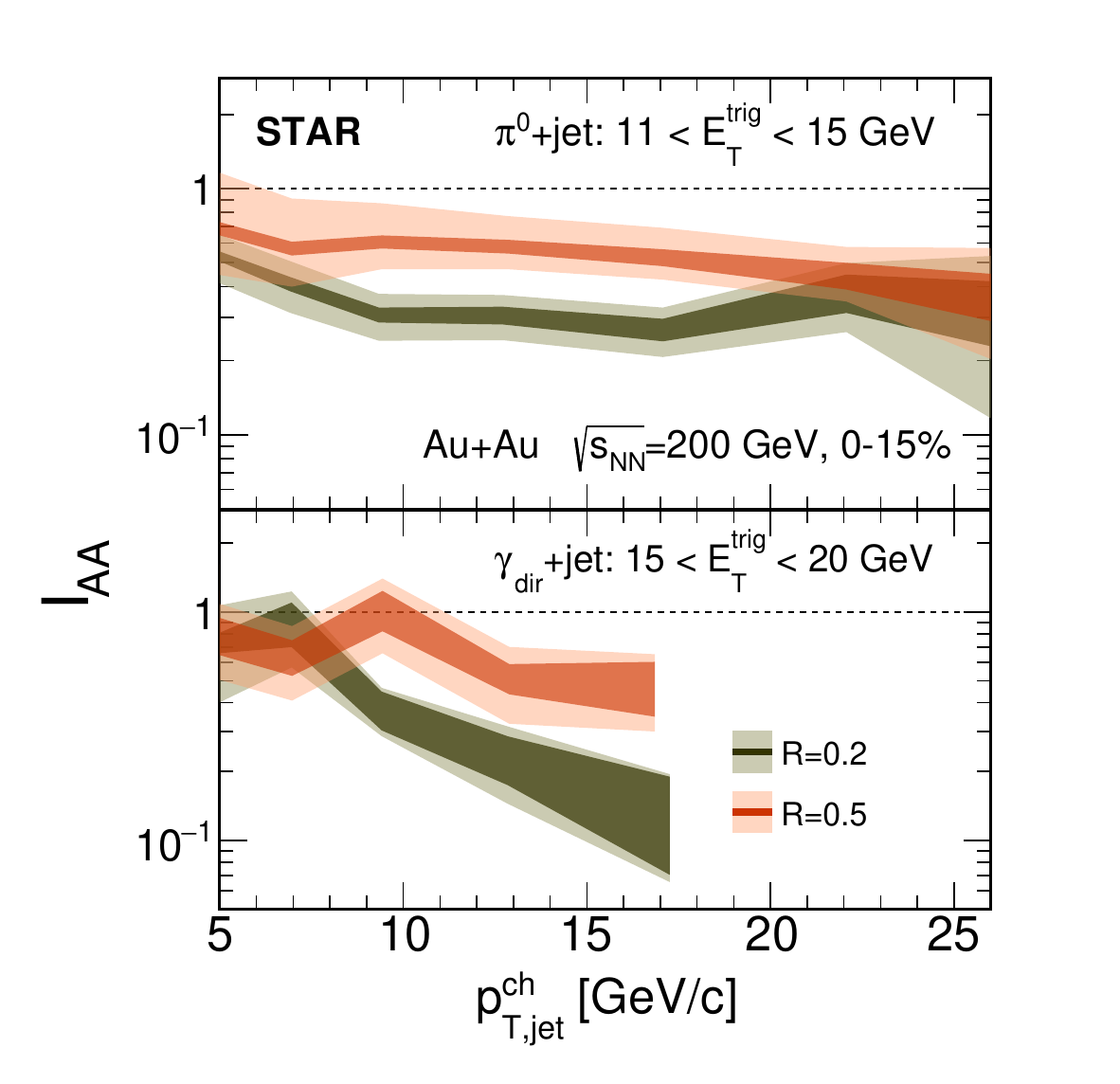}
\vspace{-2mm}
\caption{(Color online) Ratio \IAA\ of semi-inclusive recoil jet distributions in central \AuAu\ collisions~\cite{STAR:2023ksv}), for \gammadir\ and \pizero\ triggers and $\rr=0.2$ and $\rr=0.5$. Uncertainty bands account for correlated uncertainties in numerator and denominator. Dark bands are the statistical error and the light bands are the systematic uncertainty.} 
\label{Fig:IAA}
\end{figure}
\noindent

{\it Results -} Semi--inclusive recoil--jet distributions measured in \pp\  and central \AuAu\ collisions, for \gammadir\ and \pizero\ triggers and for recoil jets with $\rr=0.2$ and 0.5, are shown in Ref.~\cite{STAR:2023ksv}. These measurements extend lower in \pTjetch\ ($\sim10$ \gev), with larger \rr, than the comparable \pTjet\  range of calorimetric $\gammadir$+jet measurements of central \PbPb\ collisions at the LHC~\cite{Chatrchyan:2012gt,Sirunyan:2017qhf,Aaboud:2018anc,CMS:2018jco,ATLAS:2023iad}, and lower  in \pTjetch\ than commonly reported for jet measurements in \pp\ collisions. Such low \pTjetch\ distributions need not be interpretable perturbatively; usage of the term ``jet'' here is operational, referring only to application of the \antikT\ clustering algorithm to measure trigger-correlated recoil-energy flow within an aperture of radius $\sim\rr$. The phenomenology of these distributions, and their comparison to theoretical calculations, can elucidate their predominant underlying physical processes, both vacuum and in-medium, and encompassing both the perturbative and non-perturbative regimes.

Figure~\ref{Fig:IAA} shows \IAA, the ratio of recoil--yield distributions in central \AuAu\ and \pp\ collisions  for a common trigger and \rr, in the highest measured \ETtrig\ bin for each trigger. Marked recoil-yield suppression in central \AuAu\ collisions is observed for $\rr=0.2$, with reduced suppression for $\rr=0.5$. Yield suppression ($\IAA<1$) arises from the combined effects of spectrum shape~\cite{STAR:2023ksv} and population-averaged energy loss, indicating smaller medium-induced energy loss for $\rr=0.5$, i.e. a larger fraction of initial jet energy is captured in cone radius 0.5 than in 0.2. This observation provides a new measurement of the angular scale over which jet energy is redistributed by quenching.

Reference~\cite{STAR:2023ksv} shows that \IAA\ is consistent within uncertainties for \gammadir\ and \pizero\ triggers in a common \ETtrig\ bin, with steeper spectrum shape. This likewise indicates larger average medium-induced energy loss for the \pizero-triggered recoil jet population,  providing new constraints on the the flavor and path-length dependence of jet quenching~\cite{STAR:2023ksv}.

Figure~\ref{Fig:Rratio} shows \Rbrtwofive, the ratio of recoil-jet yields for $\rr=0.2$ and 0.5 that probes the transverse jet energy profile, for \pp\ and \AuAu\ collisions~\cite{STAR:2023ksv}. Measurements of this jet shape observable in \pp\ collisions provide precise tests of pQCD and Monte Carlo generator calculations~\cite{ALICE:2013yva,ALICE:2019qyj,ALICE:2019wqv,ALICE:2023ama,CMS:2014nvq,CMS:2020caw,Adam:2015doa,ALICE:2023jye,Soyez:2011np,Dasgupta:2016bnd}. The distribution  for \pp\ collisions in Fig.~\ref{Fig:Rratio}  is seen to be well-described for both triggers by PYTHIA 6 STAR tune~\cite{Sjostrand:2006za,Adam:2019aml}. Marked suppression of \Rbrtwofive\ is observed for central \AuAu\ compared to \pp\ collisions; its value in the range $10<\pTjetch<15$ \gev\ is $0.26\pm0.09$ (sys) for \AuAu\ and  $0.50\pm0.06$ (sys) for \pp\ collisions, with statistical error smaller than the systematic uncertainty. This observation is significant evidence of medium-induced intra-jet broadening. 

Medium-induced intra-jet broadening and energy recovery at angular scale $\rr\sim0.4$ was observed in central \AuAu\ collisions at RHIC for a jet population biased towards fragmentation into high-\pT\ constituents~\cite{STAR:2016dfv}, qualitatively similar to Fig.~\ref{Fig:Rratio}. Medium-induced jet broadening has also been observed in central \PbPb\ collisions at the LHC for higher \pTjet\ and smaller angular range ($\rr<0.3$) than this measurement, with an observable that is sensitive to jet shape but not yield suppression~\cite{CMS:2018jco}. In contrast, jet substructure measurements indicate jet narrowing due to quenching~\cite{ALICE:2018dxf,ALICE:2021mqf,ATLAS:2022vii,ALICE:2023dwg,ATLAS:2023hso}, which has been attributed to selection bias~\cite{Casalderrey-Solana:2016jvj,Brewer:2021hmh}. A comprehensive understanding of these diverse observations requires their consistent modeling within a single theoretical framework.

Towards that goal, we compare the measurements in Figs.~\ref{Fig:IAA} and~\ref{Fig:Rratio} to current model calculations incorporating jet quenching (see also Ref.~\cite{STAR:2023ksv}): Jet-fluid model~\cite{Chang:2016gjp}; Linear Boltzmann Transport (LBT) model~\cite{Luo:2018pto}; Coupled linear Boltzmann transport and hydro model (CoLBT-hydro)~\cite{Zhao:2021vmu,Chen:2017zte}; Soft Collinear Effective Theory (SCET) model~\cite{Sievert:2019cwq}; and Hybrid model~\cite{Casalderrey-Solana:2018wrw}.  All models utilize PYTHIA~\cite{Sjostrand:2006za} for initial conditions, hard-process generation, and vacuum jet--shower evolution. Figure~\ref{Fig:Rratio} and Ref.~\cite{STAR:2023ksv} show that the \pp\ baseline calculation for these jet quenching models is consistent with the current measurements.

For modeling the QGP in \aaa\ collisions, a significant difference between the calculations is their treatment of ``back-reaction,'' which is excitation of the QGP medium by jet-medium interactions. Back-reaction generates a jet-associated wake, which is correlated with the jet and included in its measurement. The LBT, CoLBT, and Hybrid models have different back-reaction implementations~\cite{STAR:2023ksv}. Comparison is made to Hybrid model calculations both with and without the wake component (Hybrid/wake and Hybrid/nowake). The jet-fluid and SCET models do not incorporate back-reaction.

\begin{figure*}[htb!]
\centering 	 	
\includegraphics[width=0.75\textwidth]{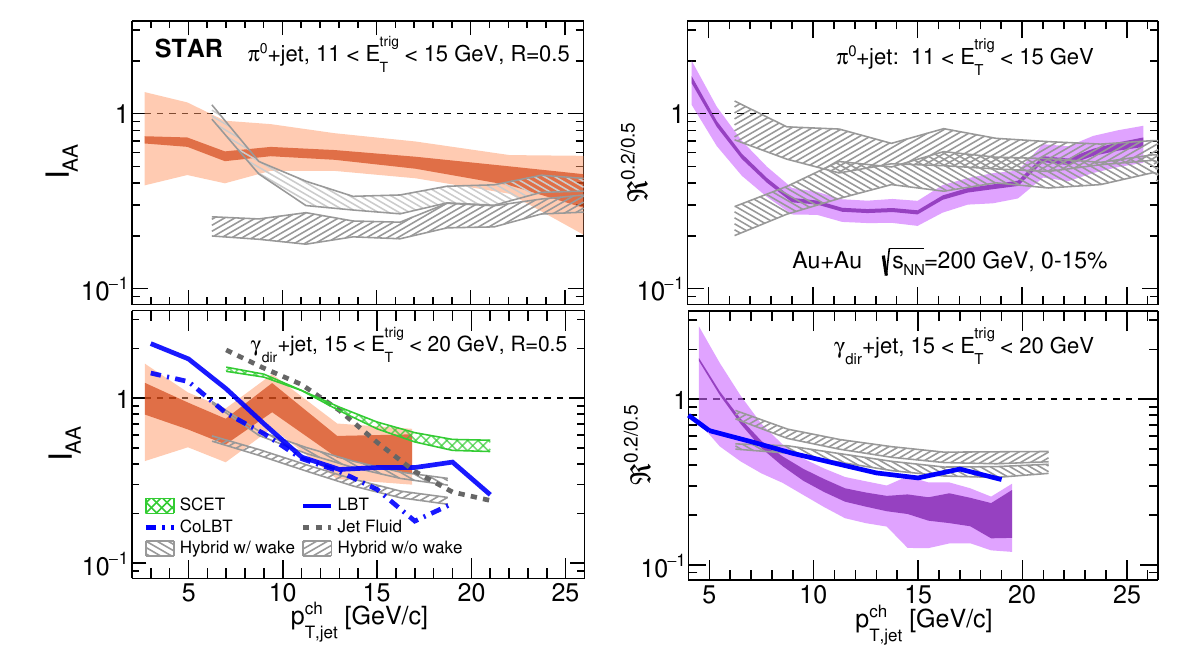}
\caption{(Color online) Comparison of model calculations to \IAA\ for $\rr=0.5$ from Fig.~\ref{Fig:IAA} (left panels) and \Rbrtwofive\ for central \AuAu\ collisions from Fig.~\ref{Fig:Rratio} (right panels). Model calculations are described in the text.} 
\label{Fig:RratioModels} 
\end{figure*} 

Figure~\ref{Fig:RratioModels} compares the model calculations to measurements of \IAA\ for $\rr=0.5$ (Fig.~\ref{Fig:IAA}) and \Rbrtwofive\ for central \AuAu\ collisions (Fig.~\ref{Fig:Rratio}). The model calculations are largely consistent within uncertainties with the \gammadir-triggered \IAA\ measurement (lower left panel). However, consistency of a model with the reported data also requires agreement with the measured \rr-dependence and with the \pizero-triggered distributions, which varies the flavor and in-medium path length distributions of the recoil jet population. A subset of model calculations is also compared to these distributions.

We first focus on the Hybrid model, whose calculations are available for all panels in Fig.~\ref{Fig:RratioModels}. Neither variant, with or without wake, agrees with the data within uncertainties over much of the measured \pTjet\ range, in each case predicting less medium-induced intra-jet broadening at angular scales $\rr\sim0.5$ than observed. Both variants of the Hybrid model likewise strongly underpredict the significant medium-induced di--jet acoplanarity broadening observed by ALICE in central \PbPb\ collisions at low \pTjet\ and large \rr~\cite{ALICE:2023qve}, which probes jet wake effects. These comparisons together indicate that this Hybrid model implementation of jet wake effects does not correctly redistribute energy withing angles $\rr\sim0.5$ with respect to the jet axis.

The LBT model likewise predicts a value of  \Rbrtwofive\  greater than that observed for the \gammadir\ trigger, consistent with the Hybrid/wake prediction. However, LBT calculations for the other distributions are not at present available. Further insight into the physical mechanisms probed by the measurements reported here require an expansion of the model calculations to incorporate \pizero\ triggers and variation in recoil jet \rr\ together with more precise measurements, in particular of \gammadir-triggered \IAA.

{\it Summary- } This Letter presents measurements of \gammadir+jet and \pizero+jet correlations in \pp\ and central \AuAu\ collisions at $\sqrtsNN=200$ over a broad range in \pTjet\ and \rr. Comparison of these measurements reveal significant medium-induced intra--jet broadening with angular scale $\rr\sim0.5$ , similar to a recent di--jet acoplanarity measurement in \PbPb\ collisions at the LHC. The predictions of leading model calculations are not fully consistent with the observations. These results represent a significant step towards understanding the physical mechanisms underlying jet interactions with Quark-Gluon Plasma.

{\it Acknowledgments- } We thank
Shanshan Cao,
Tan Luo, 
Guang-You Qin,
Abhijit Majumder,
Daniel Pablos,
Krishna Rajagopal,
Chathuranga Sirimanna,
Xin-Nian Wang,
and Ivan Vitev
for providing theoretical calculations.
We thank the RHIC Operations Group and RCF at BNL, the NERSC Center at LBNL, and the Open Science Grid consortium for providing resources and support.  This work was supported in part by the Office of Nuclear Physics within the U.S. DOE Office of Science, the U.S. National Science Foundation, National Natural Science Foundation of China, Chinese Academy of Science, the Ministry of Science and Technology of China and the Chinese Ministry of Education, the Higher Education Sprout Project by Ministry of Education at NCKU, the National Research Foundation of Korea, Czech Science Foundation and Ministry of Education, Youth and Sports of the Czech Republic, Hungarian National Research, Development and Innovation Office, New National Excellency Programme of the Hungarian Ministry of Human Capacities, Department of Atomic Energy and Department of Science and Technology of the Government of India, the National Science Centre and WUT ID-UB of Poland, the Ministry of Science, Education and Sports of the Republic of Croatia, German Bundesministerium f\"ur Bildung, Wissenschaft, Forschung and Technologie (BMBF), Helmholtz Association, Ministry of Education, Culture, Sports, Science, and Technology (MEXT)
and Japan Society for the Promotion of Science (JSPS).


\bibliographystyle{apsrev4-1}


\bibliography{references}

 
\end{document}